\documentclass{aa}
\usepackage[varg]{txfonts}
\usepackage{multirow}
\usepackage{textcomp}

\usepackage{ifpdf}
\usepackage{color}
\usepackage{graphicx}  
\usepackage{float}
\usepackage{amsmath}
\usepackage{acronym}
\usepackage{multirow}
\usepackage{xspace}
\usepackage{units}
\usepackage{aas_macros}
\usepackage{natbib, hyperref}
\usepackage{ulem}

\hypersetup{
    colorlinks=true, 
    linktoc=all,     
    linkcolor=blue,  
    citecolor=red,
}

\begin{document}

\title{Tiling strategies for optical follow-up of gravitational wave triggers by wide field of view telescopes}

\author{Shaon Ghosh\inst{\ref{inst1}, \ref{inst3}}\and Steven Bloemen\inst{\ref{inst1}, \ref{inst3}}\and Gijs Nelemans\inst{\ref{inst1}, \ref{inst3}, \ref{inst4}}\and Paul J. Groot\inst{\ref{inst1}}\and Larry R. Price\inst{\ref{inst2}}}

\institute{Department of Astrophysics, IMAPP, Radboud University, 6500 GL Nijmegen, The Netherlands\label{inst1}
\and LIGO Laboratory, California Institute of Technology, Pasadena, CA 91125, USA\label{inst2}
\and Nikhef - National Institute for Subatomic Physics, Science Park 105, 1098 XG Amsterdam, The Netherlands\label{inst3}
\and Institute of Astronomy, KU Leuven, Belgium\label{inst4}
}

\abstract {} {Binary neutron stars are among the most promising candidates for joint gravitational-wave and electromagnetic astronomy. The goal of this work is to investigate the strategy of using gravitational wave sky-localizations for binary neutron star systems, to search for electromagnetic counterparts using wide field of view optical telescopes.} 
{We examine various strategies of scanning the gravitational wave sky-localizations on the mock 2015-16 gravitational-wave events. First, we studied the performance of the sky-coverage using a naive tiling system that completely covers a given confidence interval contour using a fixed grid. Then we propose the {\it ranked-tiling} strategy where we sample the localization in discrete two dimensional intervals that are equivalent to the telescope's field of view and rank them based on their sample localizations. We then introduce an optimization of the grid by iterative sliding of the tiles. Next, we conducted tests for all the methods on a large sample of sky-localizations that are expected in the first two years of operation of the LIGO and Virgo detectors. We investigate the performance of the ranked-tiling for telescope arrays and compare their performance against monolithic giant field of view telescopes. Finally, we studied the ability of optical counterpart detection by various types of telescopes.} 
{Our analysis reveals that the ranked-tiling strategy improves the localization coverage over the contour-covering method. The improvement is more significant for larger field of view telescopes. We also find that while optimization of position of tiles significantly improve the coverage from contour-covering tiles,  the same for ranked-tiles leads to negligible improvement in coverage of the sky-localizations. We observed that distributing the field of view of the telescopes into arrays of multiple telescopes significantly improves the coverage efficiency by as much as $50\%$ over a single large FOV telescope in 2016 localizations while scanning $\sim 100$ deg$^2$. Finally, in our analysis for a range of wide field-of-view telescopes we found improvement of counterpart detection upon sacrificing coverage of localization in order to achieve greater observation depth for very large field-of-view - small aperture telescopes, especially if the intrinsic brightness of the optical counterparts are weak.} {}


\maketitle

\section{Introduction}
\label{sec:intro}

The discovery of the gravitational-waves (GW) event GW150914 from a coalescence of a binary stellar-mass black hole system by the Laser Interferometer Gravitational-wave Observatory (LIGO)  has established our ability to detect and measure perturbations of spacetime due to events of astrophysical origin on earth based detector. \citep{PhysRevLett.116.061102}. 
Further upgrades of the LIGO \citep{0264-9381-32-7-074001} will improve its sensitivity by a factor of few resulting in an increase in the detection volume. The advanced Virgo \citep{0264-9381-32-2-024001} scheduled to come online later this year will add to the network a third kilometer-scale detector that would improve the sky-localization of sources. As a consequence of these two developments we will be able to conduct gravitational-wave astronomy for the first time.
The estimated rates of double neutron star binary mergers, both from theoretical estimates as well as extrapolations based on the known sample of binary radio pulsars, suggest a detection rate of several tens per year for the
final advanced LIGO-Virgo network at their design sensitivity \citep{Abadie:2010cf}. The detection rate of neutron star -- black hole mergers is based solely on theoretical estimates and could well be similar in part due to a larger detection horizon at a given detector sensitivity.

Compact binary coalescences (CBC) are extremely energetic event which may also provide us an electromagnetic (EM) counterpart if one of the binary components is a neutron star \citep{Eichler:1989fk, Narayan:1992iy, 1538-4357-507-1-L59}. In case both EM as well as GW data are available the scientific yield of the detection will be significantly enhanced: the EM counterpart will for
instance provide a very accurate position and possibly redshift.
Additional information on the type of object may also be obtained, for example EM signals can help us distinguish between neutron star - black hole binaries with a massive neutron star from binary black hole systems. One can also expect to understand the physics of the merger better as encoded in the properties
of the ejecta and an estimate of the orbital inclination can be more accurately constrained \citep{1538-4357-487-1-L1}. The accurate position can
not only be used to aid the GW data analysis, but also allows a detailed study of the environment of the merger, which in turn could provide crucial information about the pre-merger evolution of the system (circumstellar material, host galaxy information, position in/outside galaxy etc).

One of the most promising candidate for joint observation of EM counterpart of a GW signal is a kilonova \citep{1538-4357-507-1-L59, Kulkarni:2005jw, Metzger21082010, 0004-637X-775-1-18, Rosswog21032014}. A kilonova is an optical/infrared signal generated from the radioactive decay of small amounts  $(\sim 0.01 M_{\odot})$ of high-angular momentum neutron star material that are ejected in the merger of a binary neutron star or a neutron star black hole binary. Kilonovae are expected to be emitted isotropically, although a slight polar dependence may be present and potentially can be used to constrain the orbital inclination \citep{Kasen21062015, Grossman21032014}. Current models indicate that the emission from kilonovae is weaker than supernovae. Assuming heavy element r-process nucleosynthesis, the peak bolometric luminosity for these events is $\sim 10^{40.5}$ -$10^{41.5}$ erg/s. This corresponds to an absolute magnitude of $-12$ to $-15$ in the optical Sloan $i$ band \citep{0004-637X-775-2-113, 0004-637X-775-1-18, Grossman21032014}. Currently the best hope to observe kilonovae is to receive external triggers from other observatories and conduct a targeted search around the triggered sky position. This is precisely the method that led to the discovery of the first kilonova by Hubble \citep{kilonova, BergerRProcess}, which was triggered by GRB 130603B. The constrained beaming angles of GRBs imply that most of these triggers for kilonovae would be located at very large distances. That is why the apparent magnitude of this event in the near-infra-red and optical was $\sim 25 - 28$. Therefore, detecting kilonova events at their typical GRB triggered distances will be a challenge for most telescopes. However, due to the isotropic nature of GW emission, one can expect to get triggers for kilonova events associated with binary neutron star coalescences at closer distances than what has been observed for the short GRB triggered kilonova. A kilonova with absolute magnitudes in the aforementioned range, within a typical LIGO-Virgo BNS detectable distance of $\sim 200$ Mpc, will have an apparent $i$ band magnitude of $\sim 21.5 - 24.5$. Furthermore, the longer duration of these events ($\sim$ days) provides us the opportunity for detailed follow-up with photometric and possibly, spectroscopic tools. These properties, namely temporal coincidence with the coalescence, isotropy of emission and long duration make kilonovae ideal candidates for EM follow-up of GW events. \\

{\noindent This brings us to the main challenges of EM follow-up of GW triggers:}
\begin{enumerate}
\item{Rapid detection and sky localization based on the GW data,}
\item{An operational set-up in which EM facilities with sufficiently large field of view (FOV) and sensitivity can react quickly to survey the GW sky-localizations, and}
\item{An efficient selection scheme in which the (candidate) counterparts can be identified in a potential sea of false positives.}
\end{enumerate}

The key element in EM follow-up of gravitational wave triggers is that we are able to detect the gravitational wave events in real time. If the gravitational wave events have associated optical counterparts, then these could last for hours to days \citep{1538-4357-507-1-L59, Berger20111, kilonova, BergerRProcess}. During the first few hours to about a day we expect the optical/infrared luminosity from these sources to be at their highest. The time and sky localization of a GW candidate are known within a minute or two after the merger signal has passed the detectors \citep{first2years, 0004-637X-804-2-114}. {\bf The GW sky-localizations will typically be as large as hundreds of square degrees \citep{lrr-2016-1, first2years}}. Following up such wide sky localization patches, within this time window, as deep as 22nd-23rd magnitude in at least two bands is a challenging task. 

Once the GW sky-localization is known the task for the optical telescopes would be to observe that area with the minimum number of telescope pointings. Since a telescope pointing would cover a {\it tile} in the sky commensurate to its field of view, the observation of any desired confidence interval of the sky-localizations would require generating a set of tiles that most efficiently captures the confidence interval. This act of capturing the confidence interval with the telescope tiles will be termed {\it coverage} in this study (Sect. \ref{Sec:skyPointing}). In this paper we discuss and compare, for the first time, the various strategies (Sect. \ref{Sec:skyPointing}) that one can implement for generating the sky-tiles for observing the GW sky-localizations using wide FOV telescopes. Based on the analysis conducted on the sky-localizations simulated by \citet{first2years}, we make recommendations on the optimal tiling strategy taking into account key aspects such as the number of telescope pointings (Sect. \ref{Sec:TileReduction}), issues of image subtraction (Sect. \ref{Sec:freeGrid}) and false-positive probabilities (Sect. \ref{Sec:FalsePositive}).

We then use the optimal tiling strategy to investigate how to optimize coverage of the sky-localization regions for a given observation area (total sky-area covered by the tiles) (Sect. \ref{Sec:DistributedFOV}). Here the notion of a distributed FOV array will be introduced and the performance in covering a population of simulated GW sky-localizations with that of traditional monolithic FOV telescopes is compared. A distributed FOV array could simply be a group of smaller wide FOV telescopes operating in a coordinated fashion from different geographical location. Finally, using the tiling strategy developed, we study the observing strategy for EM counterpart of GW events (Sect. \ref{Sec:DepthCoverage}). Here we analyze how to optimize the depth of the observation, tuning it against the coverage of the GW localization. 

\section{Sky-tiling for gravitational wave localization}
\label{Sec:skyPointing}

The rapid gravitational wave sky-localization algorithm, BAYESTAR  \citep{first2years, Singer:2015xha} pixelates the sky using the HEALPix (Hierarchical Equal Area isoLatitude Pixelization) projection, computes the sky localization probability distribution function (p.d.f) of the GW event at every pixel, and then outputs this information in FITS file format. Let us define this p.d.f as a function $L(\alpha, \delta)$, where $\alpha$ and $\delta$ are the sky coordinates. We define $S_{95}$ as the surface on the sky with the smallest possible area that contains $95\%$ of the total localization p.d.f \citep{PhysRevD.89.084060} \footnote{The $95\%$ localization probability is chosen as an example for the purpose of illustration.}
\begin{equation} 
\begin{aligned}
\label{eq:LocContour}
\iint \limits_{S_{95}} L(\alpha, \delta) \,{\mathrm d} \Omega = 0.95\,,
\end{aligned}
\end{equation}
where $d\Omega$ is the solid angle subtended by an infinitesimal surface element on the celestial sphere at the center of the earth. Any telescope observing this confidence interval of GW sky-localization can choose the minimum number of telescope pointings required to enclose $S_{95}$. For every such area $S_{95}$ there exists (at least) an area $\tau_{95} (\geq S_{95})$ that is constructed out of the tiles needed for covering it. The tiles enclosing $S_{95}$ forms a subset of the tiles that cover the full sky defining a {\it sky-grid}. The $N_{CC}$ tiles constituting this area, $\tau_{95}$, are the required sky-tiles that enclose the $95\%$ confidence interval and we call them {\it contour-covering tiles} or CC-tiles. This is the most straightforward and simple tiling strategy.

If the coordinates of the grid are not fixed on the plane of the sky ({\it free-grid}), then the area $\tau_{95}$ can be obtained by optimizing the grid location in the sky. On the other hand, if the grid is fixed on the sky ({\it fixed-grid}), then $\tau_{95}$ is uniquely defined. Many optical surveys, including the currently active PTF \citep{PTFRau, PTFLaw} and Skymapper projects \citep{2007PASA...24....1K} and the future BlackGEM \footnote{https://astro.ru.nl/blackgem/} and ZTF facilities \citep{ZTF, ZTFBellm, ZTFSmith}, use a fixed grid on the sky. This grid is predefined and covers the whole sky that is visible from the observatory's location. Although a more flexible grid will in general lead to more efficient contour coverage, a fixed grid approach has advantages. Firstly, the search for optical transients is typically performed by subtracting a reference image taken at an earlier epoch, and looking at the residual image to find new sources. Distortions caused by optical elements in the telescope and the camera vary with the position in the FOV of the telescope. Looking at the same part of the sky on roughly the same position in the FOV all the time limits the complexity of image resampling. The more complex this process, the more artifacts would be present in the residuals, which can be picked up as false-positive detections. Finally, a fixed grid simplifies the data flow, storage and access, since the images taken at a certain sky coordinate of interest can easily be retrieved based on the ID of the field that contains the coordinate. 

The area $S_{95}$ defined in Eq. \ref{eq:LocContour} is unique for a given event. For an EM observer what is more important however is the $\tau_{95}$ and not the $S_{95}$. Minimizing the $\tau_{95}$ will result in the most effective tiling strategy. Instead of choosing the minimum number of tiles required to enclose the smallest confidence interval contour, we can sample the entire sky-localization map with discrete 2D-intervals equal to the FOV of the telescope and select the smallest number of these sampled intervals that constitutes $95\%$ localization posterior probability.

Consider a telescope with a FOV of $\Delta \alpha \Delta \delta$ for which we would want to construct the sky-tiles required for observation. We can construct an equal area grid on the sky with grid spacing $\Delta \alpha$ and $\Delta \delta$ along the right ascension and declination respectively to cover the entire sky with an integral number of tiles. The probability of localization at an arbitrary sample on the grid would thus be given by
\begin{equation} 
\label{Eq:tileValue}
\begin{aligned}
T_{ij} = \int_{\alpha_i}^{\alpha_i + \Delta \alpha} \int_{\delta_j}^{\delta_j + \Delta \delta} L(\alpha, \delta) \,{\mathrm d} \Omega\,,
\end{aligned}
\end{equation}
where, $(i,j)$ are the coordinates of the sample in the sky-localization map. After ranking based on the value of the integral in Eq. \ref{Eq:tileValue} we can select from the top of the list of these samples until we have reached a cumulative probability $\geq 95\%$. We call these sampled intervals of sky-localizations the {\it ranked-tiles} (RT). It is straightforward to show that this strategy is guaranteed to give us a number of tiles $N_{RT}$ that is less than or equal to $N_{CC}$ (App. \ref{app:proof}).

\begin{figure}[t]
\centering
\includegraphics[width=6.0cm]{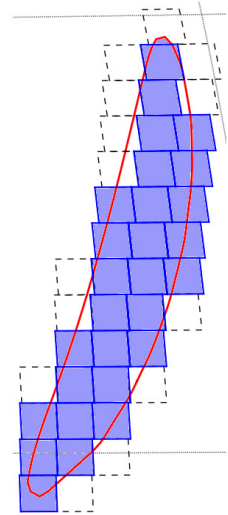}
\caption{Comparison between tiling generated for a three detector network for the two tiling strategies. The contour shows the smallest $95\%$ credible area on the sky as obtained from BAYESTAR. The tiles (both dashed and solid lines) constitutes the tiles required to cover this contour (CC-tiles). This set of tile contains $96.5\%$ localization likelihood. Shaded tiles are the ones that we obtain from the ranked-tiles. We note that less number of ranked-tiles are required to cover the $95\%$ localization than what we would require if we were to cover the contour.}
\label{fig:3DetComp}
\end{figure}

In Fig. \ref{fig:3DetComp} we show the CC-tiles and the ranked-tiles for a gravitational-wave sky-localization observed by a wide field telescope with FOV = 2.7 deg$^2$. The ranked-tiles are shown by blue solid tiles and the CC-tiles are all the tiles including the blue solid tiles and the dashed unfilled tiles. The $S_{95}$ surface is enclosed by the red contour. In this case, the ranked-tiles are a subset of the 42 CC-tiles, reducing the number of tiles by 14. 

\section{Sky tiling in the first two years LIGO - Virgo observation scenarios}
\label{Sec:firstTwoYears}
An astrophysically motivated simulation of low latency detection and rapid sky localization for the first two years of LIGO and Virgo operation was presented in \cite{first2years}. About a hundred thousand binary neutron star sources with different intrinsic parameters were injected in simulated 2015 and 2016 detector noise power spectral density (PSD). Out of these Singer et al. detected around 1000 injections using the low latency detection pipeline \citep{svdPaper}. They localized all the detected sources using BAYESTAR. For a thorough understanding of the observing scenarios during the first two years using optical telescopes we generated tiles for localizations employing both the methods discussed earlier. First, we will present the results of the comparison between the two methods. Then we will relax the fixed-grid constraint and investigate the coverage resulting from the optimization. 

\subsection{Comparison of the ranked-tile method and the $95\%$ confidence interval tiling in the first two years}
\label{Sec:TileReduction}
\begin{figure}[t]
\centering
\includegraphics[width=8.0cm]{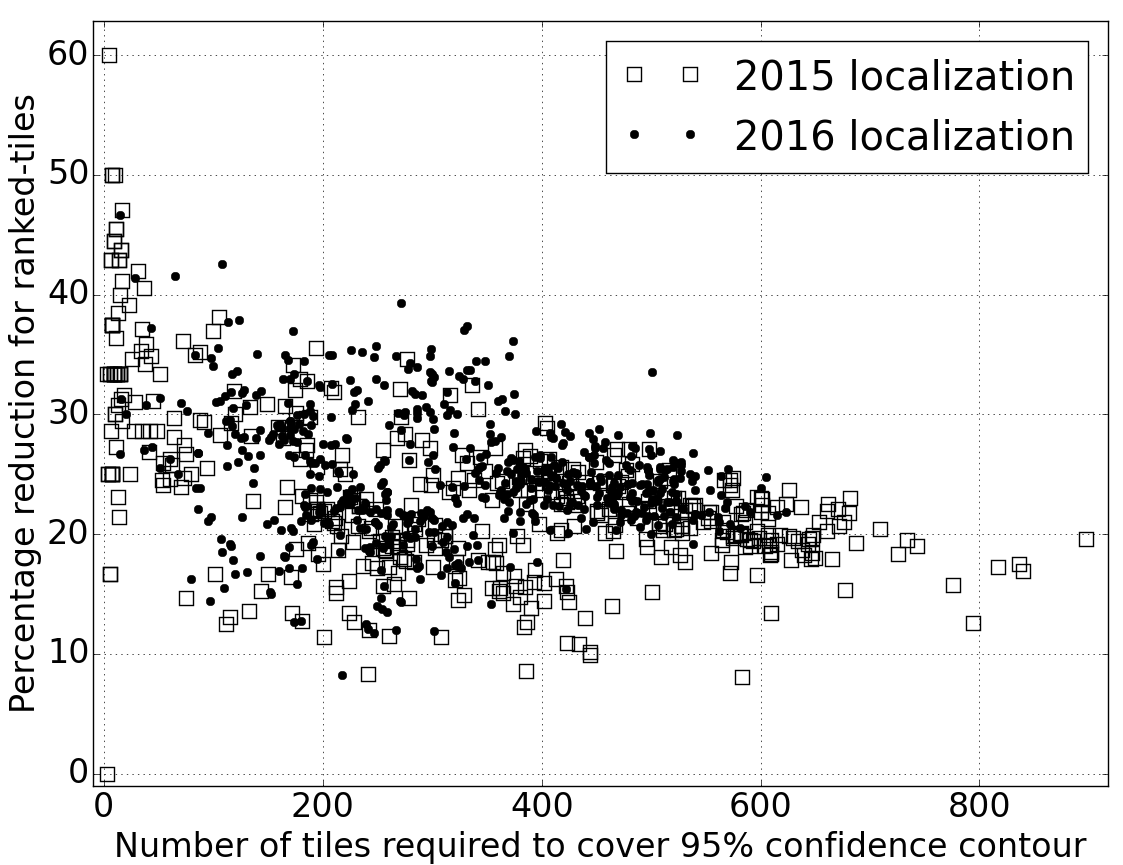}
\caption{Fractional reduction (\%) in the number of tiles required to cover 95\% GW sky-localization in 2015 and 2016 due to implementation of the ranked-tiling strategy. Note the trend of lower reduction for larger sky-localization. (The apparent bimodality in the tile reduction is accidental and disappears for other confidence intervals and FOVs. ) }
\label{fig:ReductionComparison}
\end{figure}

\begin{figure}[tb]
\centering
\includegraphics[width=8.0cm]{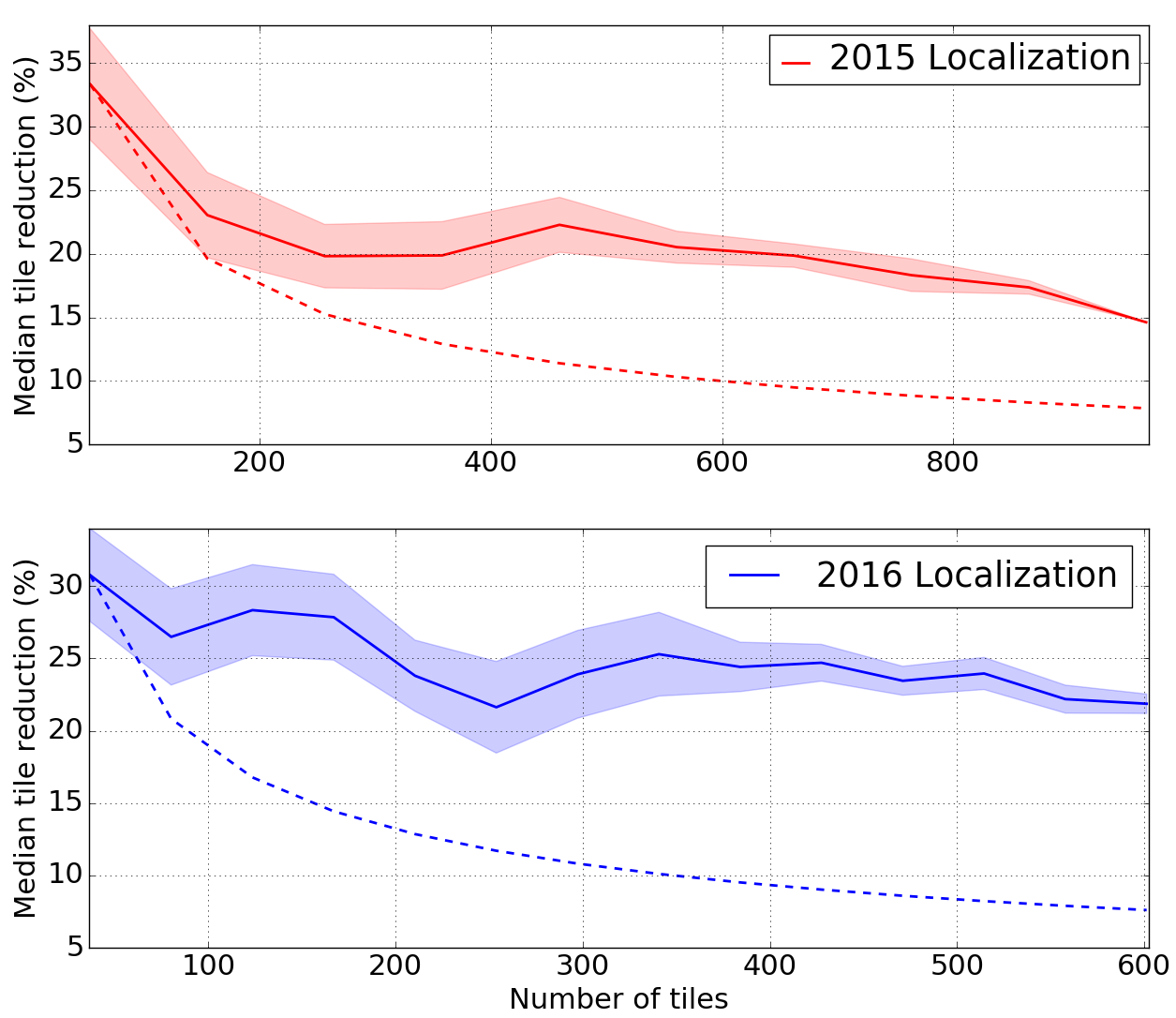}
\caption{Median reduction in the number of tiles in the various tile bins illustrating the lowering of the reduction as a function of the GW sky-localization size in 2015 and 2016. The shaded regions represent the root mean square variation of the percentage reduction of tiles in the bins. The dip near the 250 tiles is an artifact of the discrete FOV. The dashed lines represents tile reduction if ranked tiles are obtained by dropping less probable peripheral CC-tiles.}
\label{fig:3DetCompBins}
\end{figure}

We show the percentage reduction in the number of tiles required from using the ranked-tile strategy instead of the CC-tiling strategy in Fig. \ref{fig:ReductionComparison}. Once again we cover the 95\% localization as an example with a 2.7 deg$^2$ FOV telescope. The tile reduction percentage is $\frac{\Delta N}{N_{CC}}\times 100$, where $\Delta N = N_{CC} - N_{RT}$. The positive values of the reduction for all the GW events from the first two years clearly show that the ranked-tiling method minimizes the number of tiles required to observe a given confidence interval of GW sky-localization. What is not immediately evident from Fig. \ref{fig:ReductionComparison} is whether this reduction in the number of tiles is always because ranked-tiles are a subset of CC-tiles (as we observed in Fig. \ref{fig:3DetComp}).
For any given GW event the sky-localization distribution is such that the lowest probability tiles are at the periphery of the set of CC-tiles. The localization area scales as $N$, the number of tiles. If it is true that the ranked-tiles are just a subset of the CC-tiles, and that they are obtained by simply dropping the least probable CC-tiles, then one would expect that the tile-reduction must scale as $\sqrt{N}$ since the circumference scales with $\sqrt{N}$. Hence, the tile reduction percentage over the CC-tiles should scale as $1/\sqrt{N}$. In Fig. \ref{fig:3DetCompBins} we plot the bin-wise median percentage reduction as a function of number of tiles. It does show the decreasing trend of the tile reduction percentage with the total number of tiles, but much less steeply than $1/\sqrt{N}$ (shown in dashed line), or, in other words, there is considerably greater gain from using ranked-tiles for larger sky-localization than one would expect if ranked-tiles were a subset of CC-tiles. 
We found that, if we consider the same number of ranked-tiles and CC-tiles, in $92\%$ of all the GW triggers ranked-tiles enclose more localization probabilities than CC-tiles. 

\begin{figure}[tb]
\centering
\includegraphics[width=9.0cm]{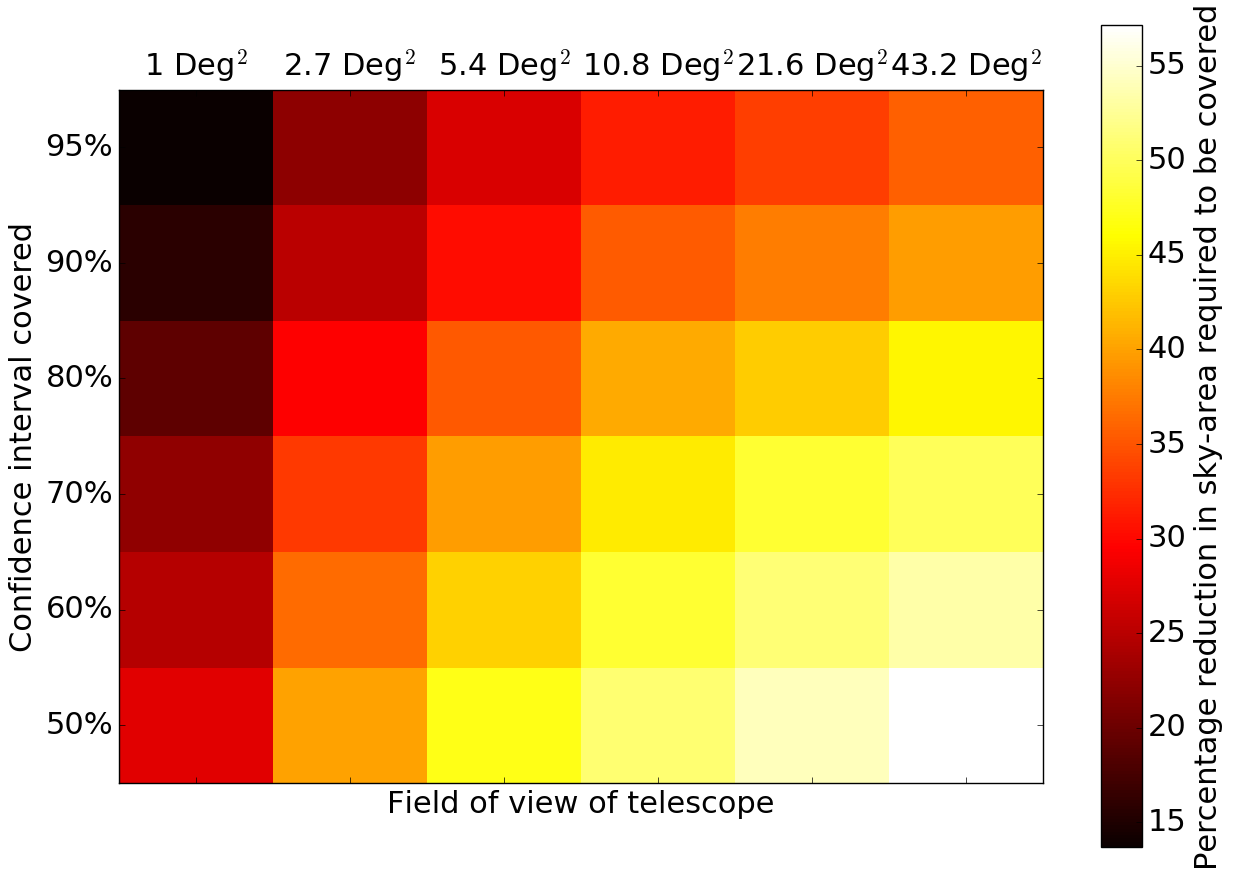}
\caption{The median reduction in sky-area required to be observed for various telescope FOVs and confidence intervals using ranked-tiles. CC-tiles are more numerous than ranked-tiles to cover the same localization likelihood. Note that since the false positive probability scales with the amount of sky-area observed, this can be interpreted as a false probability reduction as well. The reduction is greater for smaller confidence intervals and larger FOVs.}
\label{fig:gainFromRankedTiles}
\end{figure}

We repeated this exercise for telescopes with different FOVs and found that smaller FOV telescopes are more likely to have such cases where localization probability contained within ranked tiles is greater than that contained within same number of CC-tiles. Furthermore, we also found that in fair number of cases there are ranked tiles that fall completely outside the contour enclosing the smallest $95\%$ localization. Once again this happens more frequently for smaller FOV telescopes. For example our analysis has shown that out of the 475 GW sky-localizations in the 2016 era simulation, in 65 cases there are at least one ranked-tile that has fallen outside the smallest $95\%$ GW localization contour for a 2.7 deg$^2$ telescope. For a 1.0 deg$^2$ telescope this number is 283 out of 475, while for a 43.2 deg$^2$ telescope it drops to just 2. These number are far greater if we target smaller confidence intervals. Thus, we have 156, 338 and 5 ranked-tiles falling completely outside the smallest $50\%$ confidence contour for the 2.7, 1.0 and 43.2 deg$^2$ FOV telescopes respectively. In Fig. \ref{fig:gainFromRankedTiles} we show the reduction in the required area of coverage resulting from the adoption of the ranked-tiling strategy for six different FOVs (1.0, 2.7, 5.4, 10.8, 21.6 and 43.2 deg$^2$) and six different localizations confidence intervals ($50\%, 60\%, 70\%, 80\%, 90\%$ and $95\%$). We find that the reduction in required sky-area is the greatest for the largest FOV telescopes and for the smallest localization confidence regions. The reduction of the sky-area has implication on false positive probability of the search which we discuss next.

\subsection{False positives}
\label{Sec:FalsePositive}
One of the main challenges of optical follow-up of GW triggers is that the optical-sky will have a larger number of transients than in any other wavelengths, that can serve as false-positives. Extragalactic transients such as supernovae are distributed uniformly over the sky. Galactic interlopers such as M-dwarf flares, binaries that were in eclipse when the reference catalog was made, outbursting CVs (novae and dwarf novae) follow the distribution of stars in the Milky Way, i.e. higher rate of false positives closer to the Galactic Plane. However, both contour covering and ranked-tiling methods cover similar parts of the sky and hence on an event-by-event basis both methods probe the same population of interlopers. Thus, a false positive probability comparison between the methods can be made assuming that the number of false-positives per square degree of sky-area is constant within the error-box, and hence the number of false-positives is proportional to the observed area. In the preceding sections we have established that the ranked-tiling strategy allows us to reduce the number of tiles required to capture the sky-localization region of any given confidence level. Thus, the percentage decrease in false positives upon employing ranked-tiles as the observing strategy instead of the CC-tiles can be written as $100\times(N_{CC} - N_{RT})/N_{CC}$. Therefore, the color coding in Fig. \ref{fig:gainFromRankedTiles} can be interpreted as the reduction in the false positive rate. Of course if one would decide to only analyze an area containing a fixed localization percentage (e.g 95\%) that is enclosed within the tile set, the CC method would by definition cover the smallest area, ($S_{95}$) that contains that percentage and thus have the smallest number of false positives. But in practice observers will be analyzing the whole area observed by the telescope and any transient candidate found outside the $S_{95}$, but within the area defined by $N_{CC}$ tiles will also be analyzed.

\subsection{Optimization in free-grid}
\label{Sec:freeGrid}
In Sect. \ref{Sec:TileReduction} we presented the results of the two strategies to cover the GW sky-localizations using the FOV of telescopes when the sky-grid is predefined. It is expected that further minimizations of tiles can be achieved if this constraint is lifted and we are allowed to move coordinates of the tiles in the grid. The covering of the GW localizations, which are essentially irregular polygons, with the least number of square tiles is a member of a class of problems called NP-complete problems \footnote{NP stands for nondeterministic polynomial time}. Any known solution of this class of problem can be verified within polynomial time (i.e., solvable in $N^k$ steps, where $N$ is the complexity of the problem and $k$ is a non-negative integer). However, there is no known method of finding the solution from first principles. In the absence of a general recipe of finding the solution we resorted to an iterative optimization, where we iteratively shifted the tiles in the grid covering the GW sky-localization to see which configuration gives us the best result. This is done in the following two steps:
\begin{enumerate}
\item{{\bf Create the initial grid to cover the GW sky-localization:} For every sky-localization we find the smallest size of the grid that is required to cover it. The right ascension (RA) and declination (Dec) values at the center of the grid lie at the mean RA and Dec value of the localization's credible region. }
\item{{\bf Optimize rows of tiles at constant declination angle:} Once the initial grid is laid, every row in the grid is slid in the horizontal direction (along the right ascension angle) by steps of $0.1^{\circ}$ to look for the configuration that requires the lowest number of tiles in that particular row. Continuing the same process for all the rows we optimize the tiling.}
\end{enumerate}
Due to the periodic nature of the initial grid, the shifting of the rows needs to be done for one tile length. We then conducted this exercise for ranked-tiles:
\begin{enumerate}
\item{Once again we started from a primary sky-grid. We created a ranked list of $T_{ij}$ values for all the samples in this grid. This gives us the ranked-tiles for this instance of sky-grid. }
\item{Then we  slide the first row of the grid in steps of $0.1^{\circ}$, record the $T_{ij}$ values of all the tiles and create the ranked-tiles for each of these instances of the sky-grid. Note that unlike in the case of the CC-tiles, here there is no {\it a priori} way of optimizing each row of ranked-tiles but must optimize the entire grid after every iteration.}
\item{Conducting this exercise for all the rows we collect the ranked-tiles for all the instances and choose the minimum of these which should be the optimal solution.}
\end{enumerate}
In Table. \ref{tab:tilingComp} we show the results of the optimization carried out on ten randomly selected sky-localizations from the first two years simulation. Note that reasonable reduction in the required number of tiles is achieved by optimization of the CC-tiles. However, similar reduction was not observed for ranked-tiling. This indicates that the ranked-tiling strategy gives us a solution that is already close to the optimal solution. The ranked-tiles are therefore an excellent approximation of the continuum sky-localization and with virtually no need for optimization. This is an important point since it liberates the observation efficiency from the choice of the sky-grid. As we argued that many wide FOV telescopes will be using a fixed sky-grid to simplify image subtraction, the independence of the ranked-tiling strategy from the choice of the grid is a major advantage.
\begin{table*}[tbh]
\begin{center}
\caption{Comparison between tiling before and after optimization for a random selection of events from \cite{first2years}. The first column lists the sky-localization ID, the second and the third column show the number of CC-tiles required to cover the 95$\%$ localization contour (CC) for the ten events with fixed-grid and optimized-grid. The fifth and sixth columns show the same for the ranked-tiles (RT). The fourth and the seventh column shows the percentage reduction upon optimization in the number of CC-tiles and ranked-tiles respectively.}
  \begin{tabular}{ r  r  r  r  r  r  r  r  }
    \hline\hline
    \bf{ID} & \bf{CC tiles} & \bf{CC-Optimized} & \bf{Tile reduction (\%)} & & \bf{RT tiles} & \bf{RT-Optimized} & \bf{Tile reduction (\%)}  \\
    \hline
    288172 & 531 & 462 & 12.99 & & 422 & 418 & 0.95 \\
    288830 & 38 & 37 & 2.63 & & 29 & 29 & 0.0  \\
    303684 & 129 & 117 & 9.3 & & 96 & 96 & 0.0  \\ 
    313831 & 5 & 4 & 20.0 & & 3 & 3 & 0.0  \\ 
    1087 & 385 & 359 & 6.75 & & 302 & 302 & 0.0  \\ 
    468530 & 307 & 273 & 11.07 & & 217 & 213 & 1.84  \\
    588762 & 466 & 437 & 6.22 & & 365 & 364 & 0.27  \\ 
    1065078 & 264 & 237 & 10.23 & & 192 & 189 & 1.56  \\
    1027955 & 10 & 9 & 10.0 & & 9 & 9 & 0.0  \\ 
    687313 & 469 & 453 & 3.41 & & 426 & 425 & 0.23  \\
    \hline
  \end{tabular}
\label{tab:tilingComp}
\end{center}
\end{table*}

\section{Optimization of observation area - monolithic vs distributed field of view}
\label{Sec:DistributedFOV}

In order to scan the GW sky-localizations spanning over hundreds of square degrees with a reasonable chance of detecting an optical counterpart one needs wide FOV telescopes. Other telescopes may only be successful in this endeavor if they incorporate additional information like galaxy catalogs and distance localization \citep{3Destimate, Gehrels:2015uga}. Such telescopes would target galaxies within the sky-localization region to search for the counterpart and are unlikely to base their search on any sky-tiling strategy, hence in the present analysis we exclude them. A list of currently operating wide FOV telescopes that participated in the first observing run of LIGO can be obtained in \citet{Abbott:2016gcq}. Also, new facilities that could participate in the electromagnetic follow-up of GW triggers in the near future can be found in \cite{0004-637X-767-2-124}, \cite{2041-8205-789-1-L5} and \cite{Chu:2015jxa}.

\begin{figure}[tb]
\centering
\includegraphics[width=9.0cm]{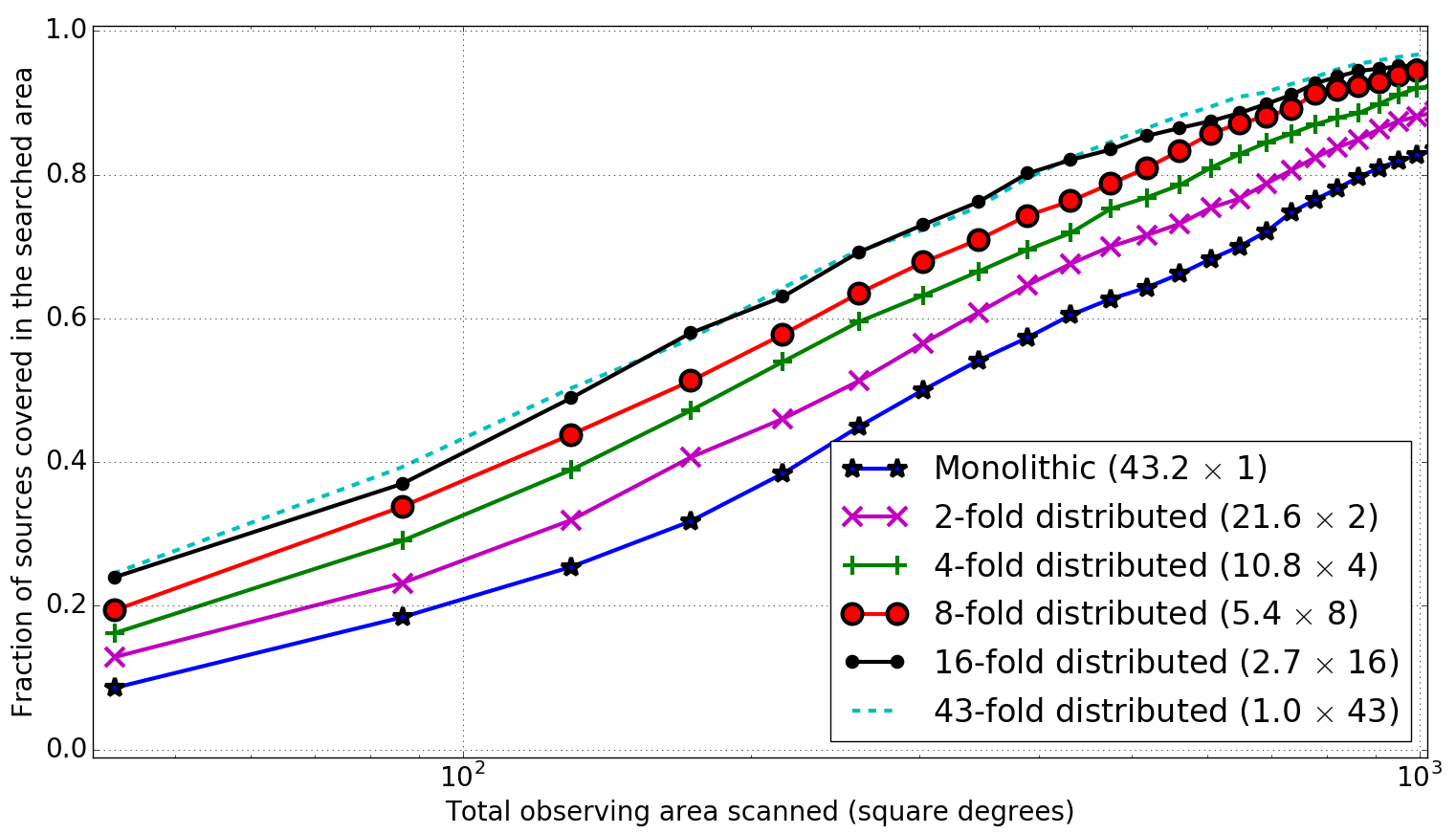}
\includegraphics[width=9.0cm]{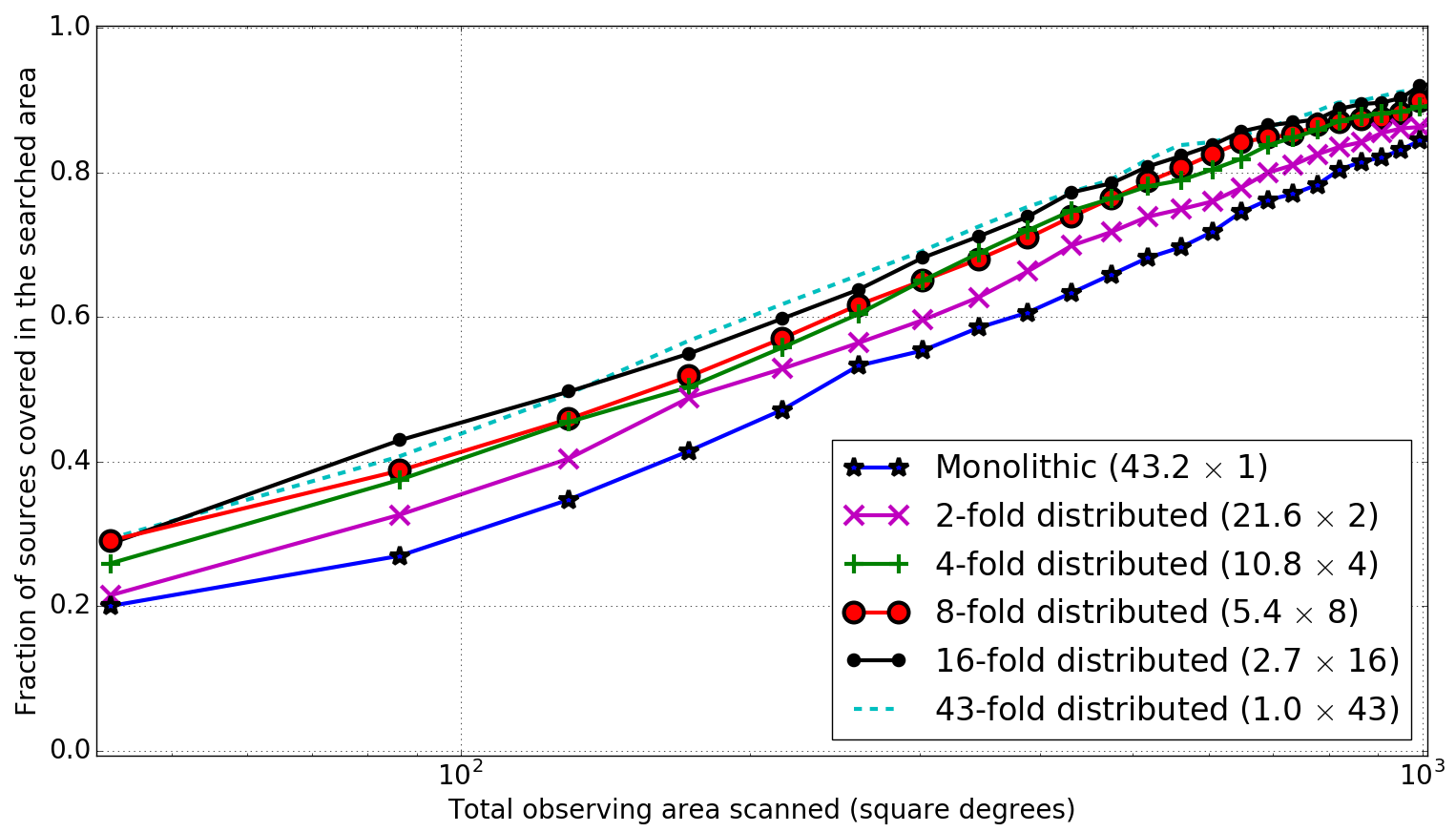}
\caption{Comparison between detection fraction as a function of sky coverage for different FOV optical observing facilities in 2015 (top) and 2016 (bottom). Here we have selected six different arrays that have the same total observing area. We see that arrays with smaller FOV in the individual telescopes are more efficient in covering the gravitational wave sky-localization.}
\label{fig:DetectionFracCompAllTelescopes}
\end{figure}

Of course the larger the FOV of a telescope, the greater is its capability to scan any given sky-localization. Although the observing area scales linearly with the FOV, the coverage of the sky-localization might scale less strongly due to the fact that above a certain size, wider angle telescopes will end up covering a lot more area extraneous to the confidence region contour than less wide angle FOV telescopes. Since smaller FOV telescopes can tile the localization contour more efficiently, one can imagine the possibility of incorporating multiple such telescopes in the form of a {\it distributed FOV array} of telescopes with a combined FOV equal to that of a large FOV telescope and expect to cover the credible region more efficiently. We performed the following analysis to test the implementation of distributed FOV arrays for ranked-tiles. Let us imagine a large FOV telescope with which we would like to scan the sky to detect the optical counterpart corresponding to the mock GW triggers from 2015 and 2016 eras. In our studies we use the largest FOV telescope from the previous analysis, namely, 43.2 deg$^2$. For each GW event we count the number of ranked-tiles that we need to observe till we reach the location of the simulated GW source. As we go down the list of the ranked-tiles and we observe a larger fraction of the sky, more event locations are covered. This is shown in the blue curves of Fig. \ref{fig:DetectionFracCompAllTelescopes} where we can see that with the increase of the total observing area the fraction of source locations that were covered increases. 
Next we distribute the 43.2 deg$^2$ FOV into two equal observing area telescopes with 21.6 deg$^2$ FOV each. Fig. \ref{fig:DetectionFracCompAllTelescopes} shows that this results in a greater fraction of coverage. Two telescopes of half the observing area are able to utilize their FOVs more efficiently to cover the highest localization regions, thereby converging to the source location faster than the original telescope with twice the FOV. Continuing this distribution of FOV further we note that the increase in detection fraction steadily increases, underscoring the benefit of using a distributed FOV array over a monolithic FOV telescope in scanning the GW sky-localizations. 

In the upper plot of Fig. \ref{fig:DetectionFracCompAllTelescopes} we see that in the synthetic 2015 sky-localization, while scanning the top 100 deg$^2$ of the ranked-tiles of the sky-localizations, the gain in the coverage of the triggers would have been $\sim 100\%$ for an array consisting of forty-three 1.0 deg$^2$ or sixteen 2.7 deg$^2$ FOV telescopes compared to a monolithic 43.2 deg$^2$ FOV telescope. The presence of the third detector in 2016 greatly improves the GW sky localizations, making the sky-localizations smaller and less elongated in general. For such localizations the coverage is less sensitive to the distribution of the FOV of the telescope. This is the reason why the improvement in the 2016 era is more modest compared to 2015 (see lower plot of Fig. \ref{fig:DetectionFracCompAllTelescopes}). Nevertheless we found a $\sim 50\%$ gain when we scan the top 100 deg$^2$ of the ranked-tiles using the smaller FOV telescope arrays (1.0 deg$^2$ and 2.7 deg$^2$ FOV) instead of a single 43.2 deg$^2$ FOV telescope. It is important to emphasize here that even though Virgo is expected to join the second observing run (O2), a significant fraction of the detections would be Hanford-Livingston double coincident events due to the combined effect of finite duty cycles and lower Virgo sensitivity \citep{Abbott:2016gcq}. Thus, the information from the localizations of the 2015 era is pertinent to the 2016 era and hence have been included in the results of this work.\footnote{Note that in 2016 the sensitivity of the LIGO detectors will also increase. The same event from 2015 will be better localized in 2016, however, it also means that the detectors will be sensitive to weaker sources that were undetectable in 2015. Thus, as a fraction of the population the localizations of the sources do not improve due to better sensitivity of 2016.}
\begin{figure}[tb]
\centering
\includegraphics[width=9.cm]{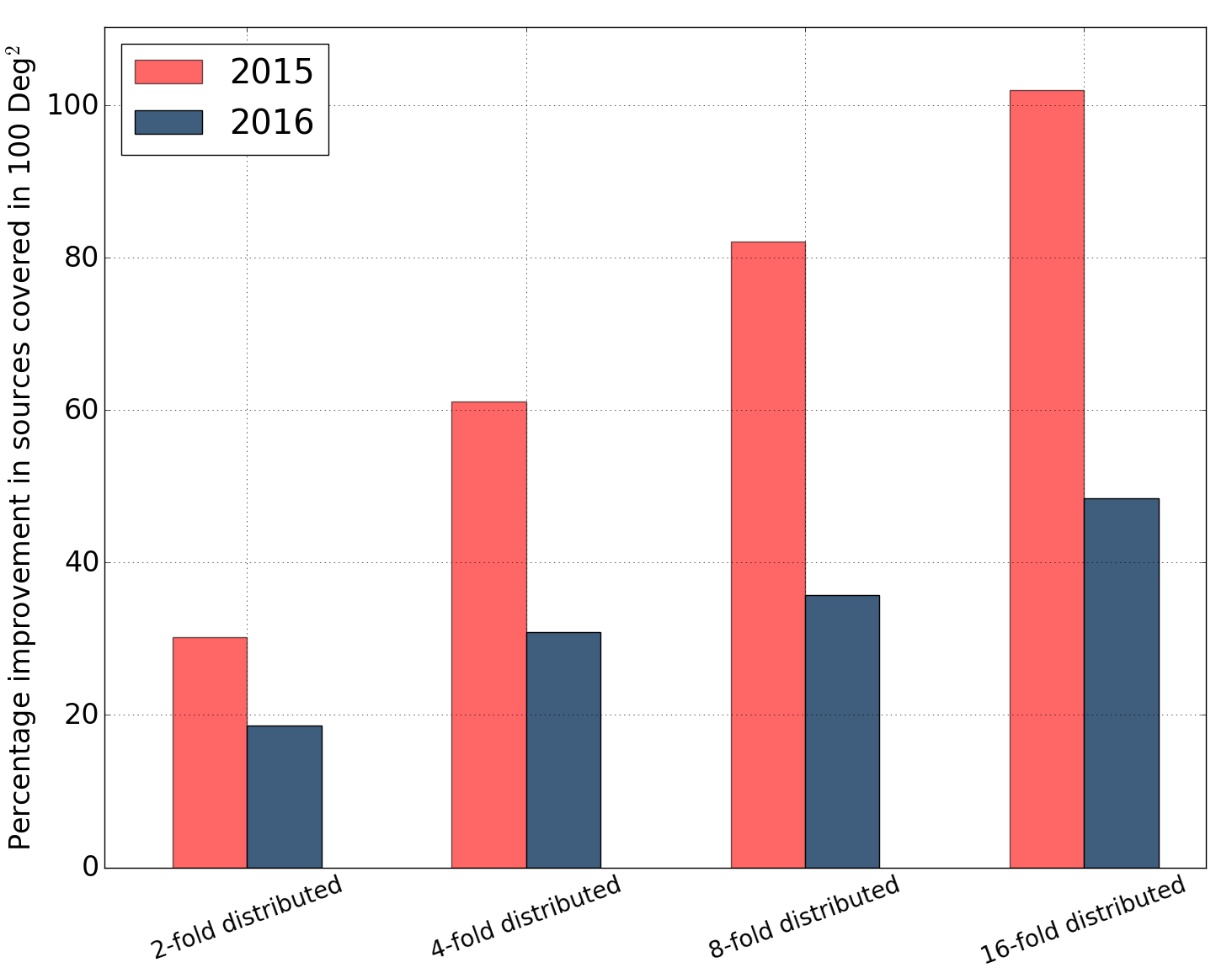}
\caption{Improvement($\%$) in the number of sources covered from using distributed FOV arrays instead of a single monolithic telescope in the highest likelihood 100 deg$^2$ sky-area. Note that in 2016 the coverage improves by as much as 50\% if we use a 16 fold distributed array instead of a single large FOV telescope. In 2015 this improvement was supposed to be even larger (as much as 100\%). This implies that the coverage improvement from using distributed FOV is even greater for double coincident events (see discussion in Sect. \ref{Sec:DistributedFOV}).}
\label{fig:DetectionFracComp100SqDeg}
\end{figure}

Note that the gain in coverage diminishes rapidly below 2.7 deg$^2$ FOV telescopes for the 2015-16 sky-localizations as is evident from the lack of any significant gain in detection upon distributing the FOV further from an array of 2.7 deg$^2$ to an array of 1.0 deg$^2$ FOV. Figure \ref{fig:DetectionFracCompAllTelescopes} also shows that the differences between the distributed FOV arrays and the monolithic FOV telescope are minimal for very large observed areas, which is to be expected since if we were to observe a very large area in the sky, we expect to cover most of the event locations regardless of the tiling strategies or whether we use a monolithic FOV telescope or a distributed FOV array. However, we will be able to observe such large areas only for very slowly varying light curves. The most interesting and meaningful part of the plots in Fig. \ref{fig:DetectionFracCompAllTelescopes} is around $\sim 100$ square degrees where distributed FOV arrays improves the coverage. This is shown in Fig. \ref{fig:DetectionFracComp100SqDeg} where we show, for the highest likelihood 100 deg$^2$, the improvement (in $\%$) of the sources covered over the single monolithic FOV telescope for the various arrays. The coverage of the source location improves by as much as $50\%$ if the observers uses distributed FOV arrays instead of a single monolithic FOV telescopes. The improvement is expected to be even better for cases where the GW events were observed by only two detectors.

\section{Depth and coverage}
\label{Sec:DepthCoverage}
\begin{figure}[tb]
\centering
\includegraphics[width=8.0cm]{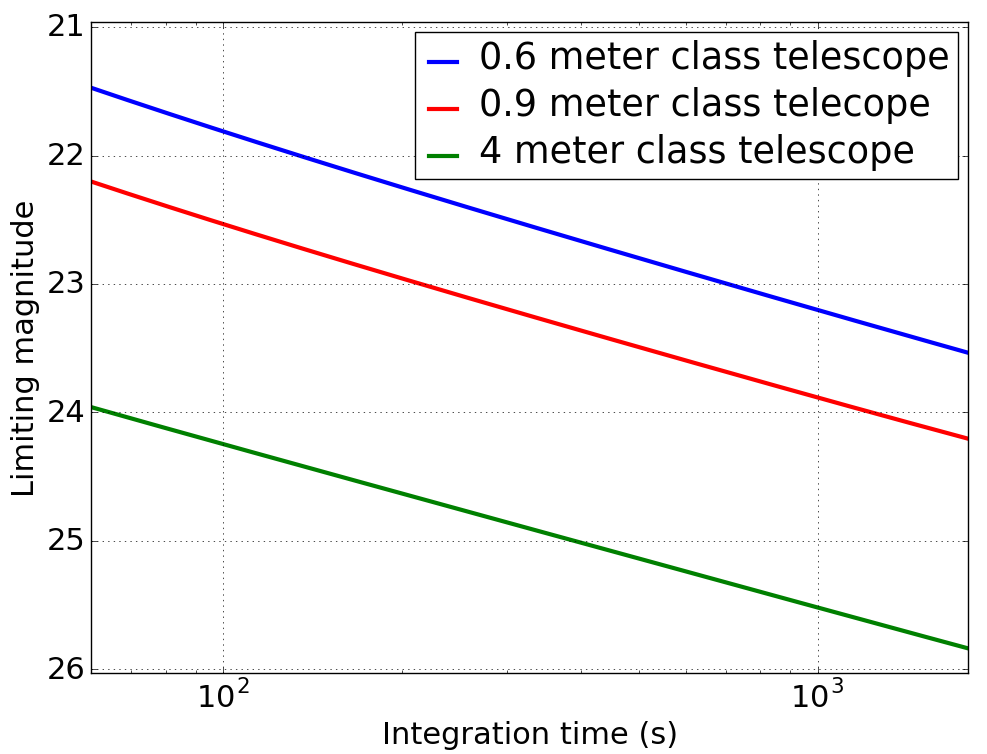}
\caption{The variation of the limiting magnitude for observation with integration time for telescopes with three different apertures.}
\label{fig:timeMag}
\end{figure}

Until now we have been discussing the tiling strategies in the context of efficient covering of the GW localization regions. However, the detection of the optical counterpart will depend on the depth of the observation and not just the mere coverage of the localization area. The depth of observation by a particular telescopes depends on various factors, including the optical seeing quality of the site, the phase of the moon on the night of observation, the air-mass of the observation, etc. However, the most important quantity is the integration time of the observation and the mirror size. Thus, for the present study we will make the simplifying assumption that all the other factors are held constant at their optimum values at a typical site (seeing $= 1.0$, airmass $= 1.0$, moon phase $= 0$ (new moon)). We present studies for three different apertures sizes, 0.6, 0.9 and 4.0 meters. The limiting magnitude as a function of the integration time is shows in Fig. \ref{fig:timeMag} \footnote{We have used the exposure time calculator from http://www.noao.edu/scope/ccdtime/ to get the limiting magnitudes for the 0.9 and 4.0 meter class telescopes and scaling them we got the same for the 0.6 meter class telescope.}. 

To make our study more general, we will not assume any geographical location of the observer. During actual observations, the GW localizations would not always be visible from the any given location. For this general study we assume that all sky-locations are entirely visible all the time. 

\subsection{Expected number of accessible sources}
\label{Sec:accesibleSources}
Increasing the integration time allows us to see deeper into the universe. If we can see deeper by a factor of $f$ then, assuming a uniform density of GW sources (which is true for non-cosmological distances), we have a factor of $f^3$ increase in the number of sources that are observable. However, larger integration time also means that in the same total observation time we will be able to observe a smaller fraction of the GW localization. Thus, the increase in depth comes at the expense of coverage. Here we will present a study in which we investigate how the number of observable sources varies with the changing integration time and coverage for a uniform volume distribution of sources if we have a total of two hours of observation time. Identification of the optical counterpart would require more than a single epoch of observation of the candidates, preferably in multiple filters to get photometric and light curve information. Thus two hours of observation per epoch is a reasonable choice. Let us assume that integrating for $t_{A}$ seconds allows us to observe at a limiting magnitude of $m_A = m(t_A)$, where $m(t_A)$ is obtained from Fig. \ref{fig:timeMag}. Similarly, integrating for $t_B$ seconds we reach a magnitude of of $m_B = m(t_B)$. If $M$ is the absolute magnitude of the source then, the ratio of the accessible distances (disregarding extinction) for the two integration times $x = D_L^A/D_L^B$ is given by, 
\begin{equation}
x = \frac{D_L^A}{D_L^B} = 10^{\frac{1}{5}[m(t_A) - m(t_B)]}\,,
\label{Eq:magnitudeExpression}
\end{equation}
where $D_L^A(D_L^B)$ are the limiting accessible distance for observations conducted with integration time of $t_A(t_B)$ seconds. Furthermore, let the total GW localization probability that the observer is able to cover if each pointing requires $t_A(t_B)$ seconds of integration be $P_A(P_B)$. Therefore, the ratio of the expected number of accessible sources for the two observations is:
\begin{equation}
\frac{n_A}{n_B} = x^3 \left(\frac{P_A}{P_B}\right)\,.
\label{Eq:factionOfSources}
\end{equation}
We present a comparative study between different depths of observation where, as a reference, we are using the $95\%$ localizations region tiling. For an event the number of ranked-tiles required to cover a particular confidence interval gives us the integration time for each pointing. Using this time and Fig. \ref{fig:timeMag} we compute the value of $x$. Thus, the integration time in the event-by-event basis is not the same, however the covered GW localization likelihood is constant. 
\begin{figure}[tb]
\centering
\includegraphics[width=8.0cm]{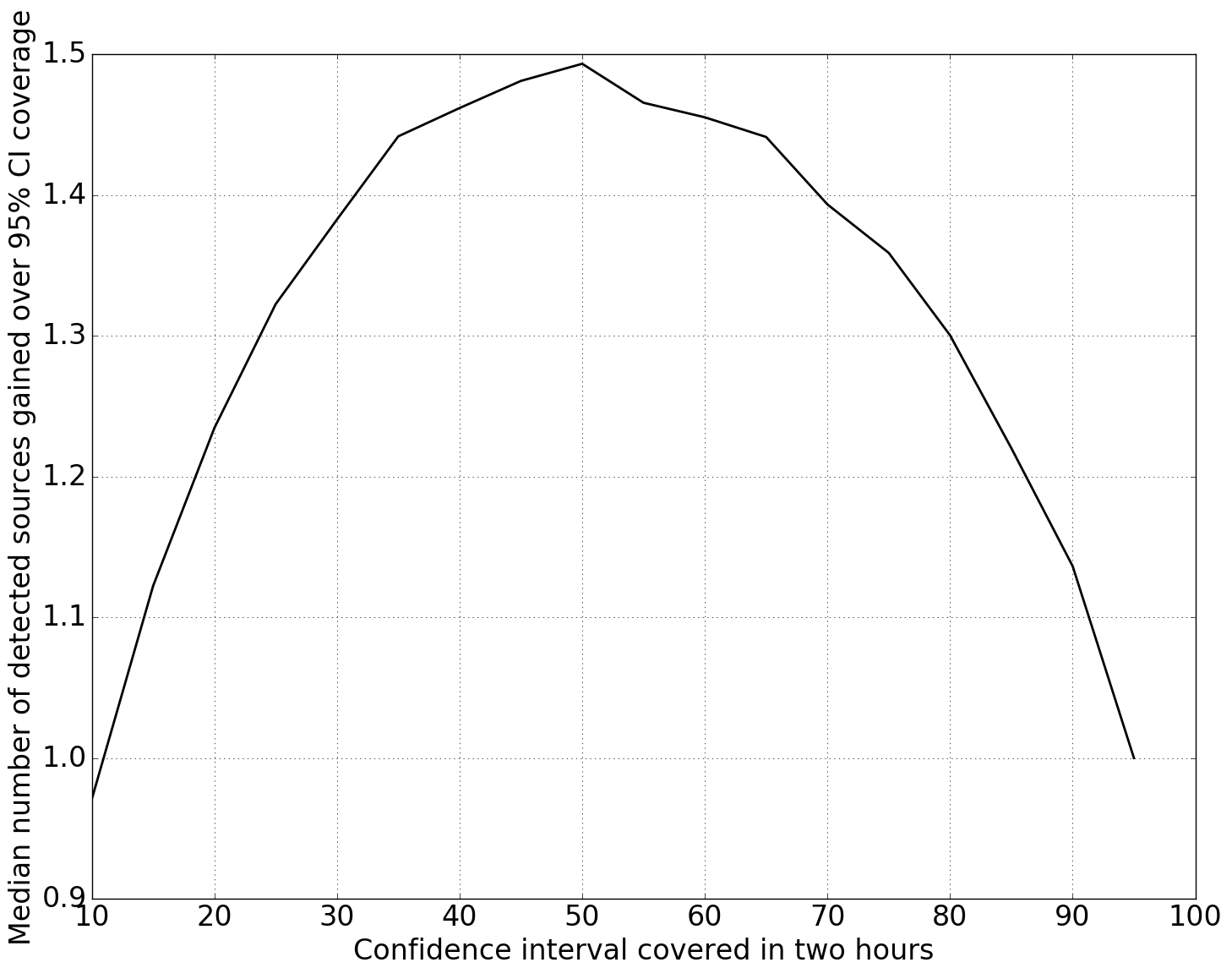}
\caption{The median of the ratio of accessible source population for various confidence intervals w.r.t accessible source population for $95\%$ confidence interval. Note that the expected number of detection for uniform volume distribution of sources peaks at an intermediate localization confidence interval. This shows the likelihood of the detection of the optical counterparts is more if the observer goes deeper at the expense of coverage of the localization. }
\label{fig:RatioOfInaccessible}
\end{figure}
We note from Fig. \ref{fig:RatioOfInaccessible} that the competing effects of larger integration time and reduced coverage result in a maximum in the expected number of accessible sources at localizations likelihood $\sim50\%$. Observing deeper gives access to a greater number of sources than covering larger localization regions. 

\subsection{Number of detectable optical counterparts of binary neutron star coalescences}
\label{Sec:ActualSources}
\begin{figure}[tb]
\centering
\includegraphics[width=8.0cm]{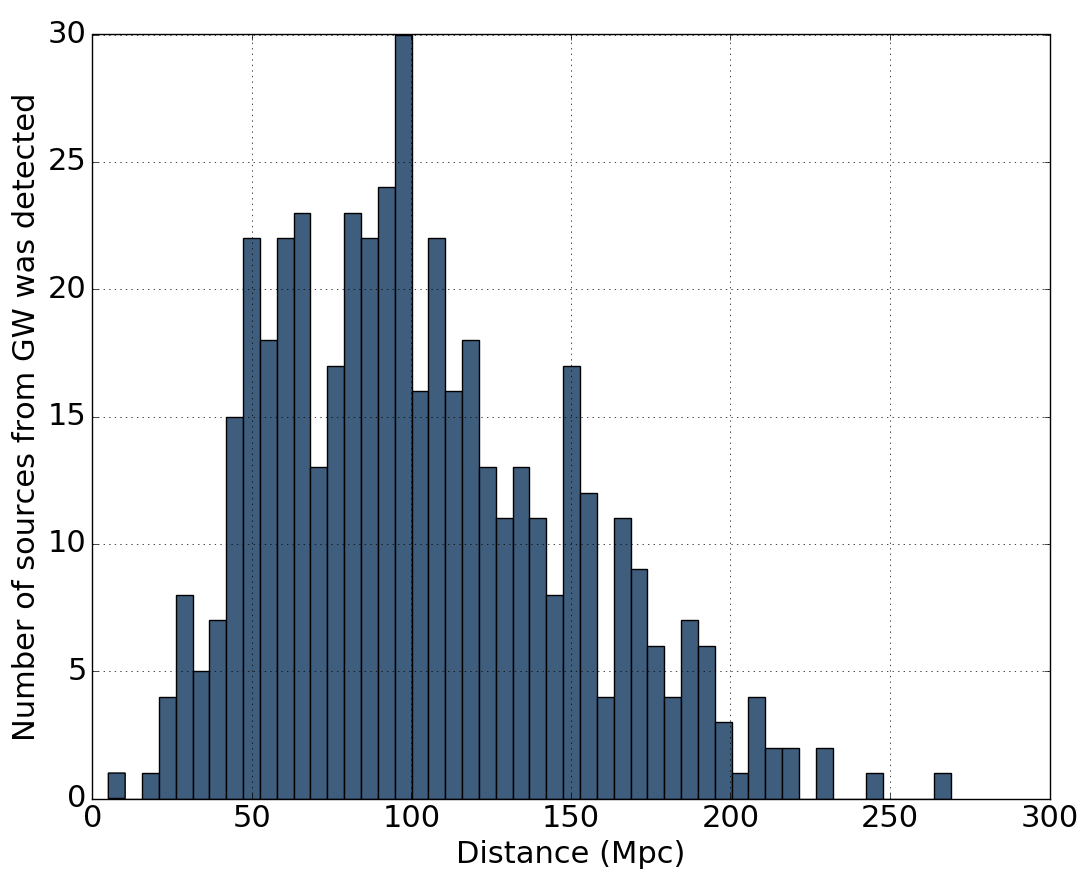}
\caption{Simulation of distance distribution of the detected events from 2016. }
\label{fig:DistanceDistribution}
\end{figure}

\begin{figure}[tb]
\centering
\includegraphics[width=7.55cm]{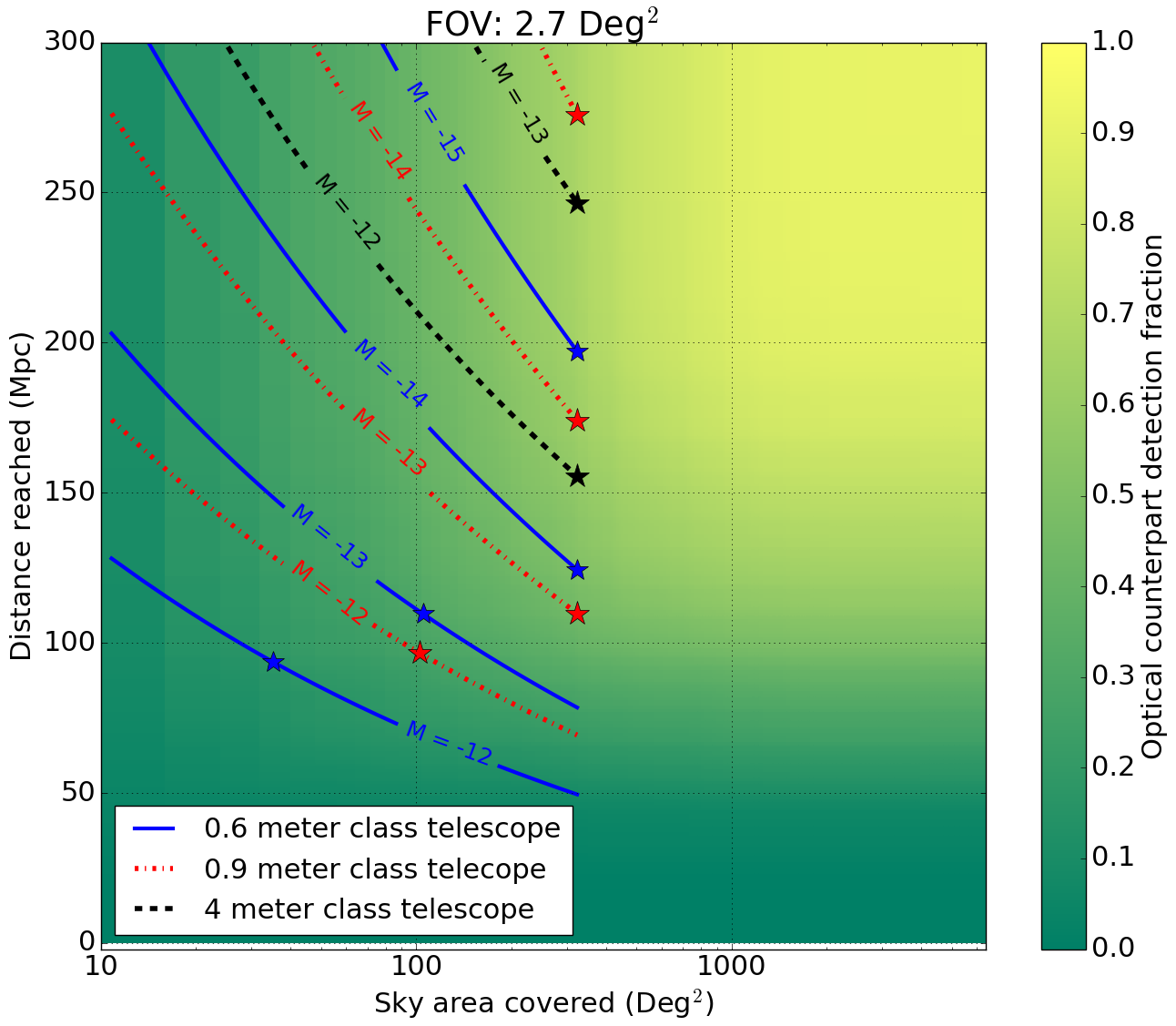}
\includegraphics[width=7.55cm]{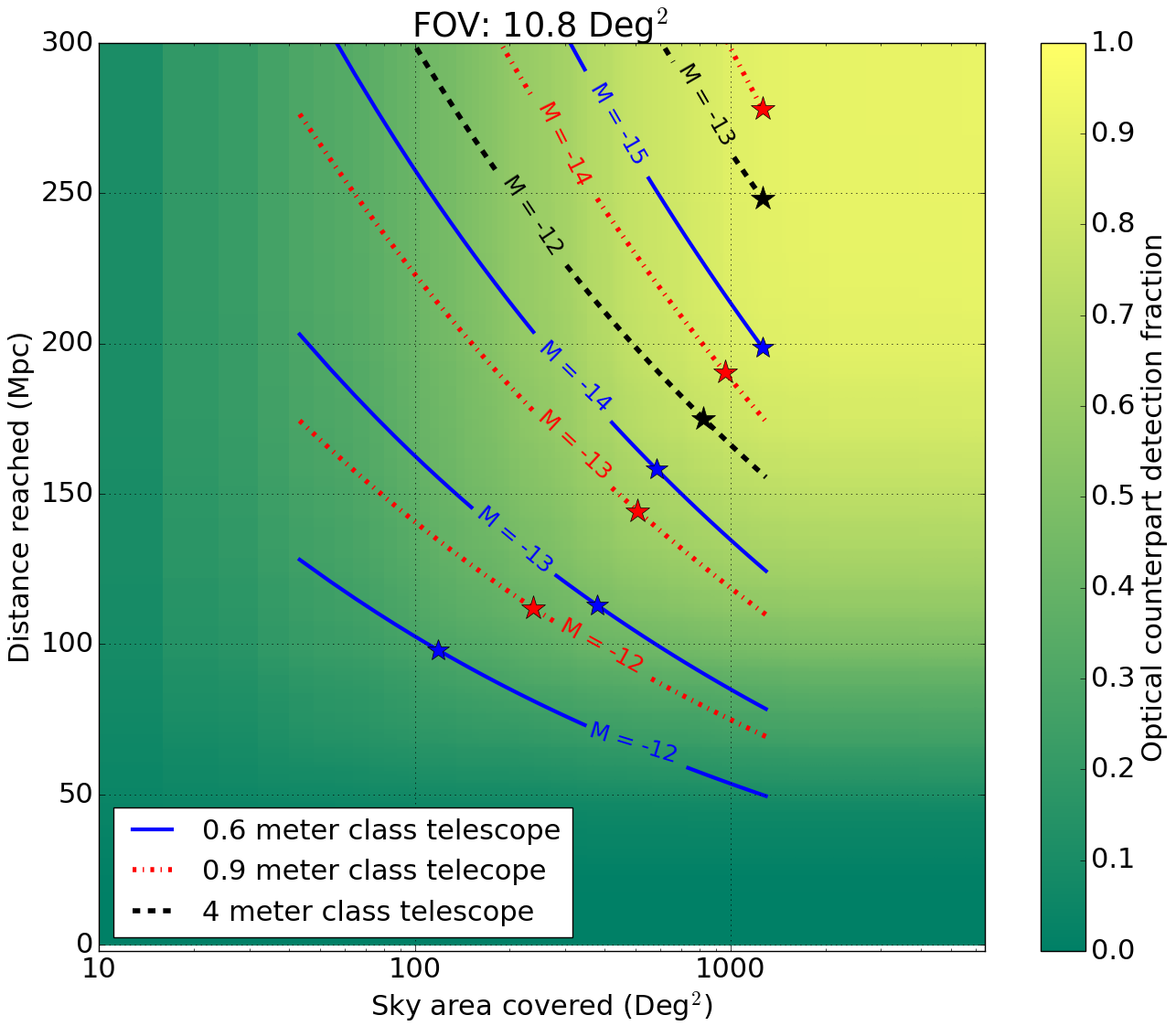}
\includegraphics[width=7.55cm]{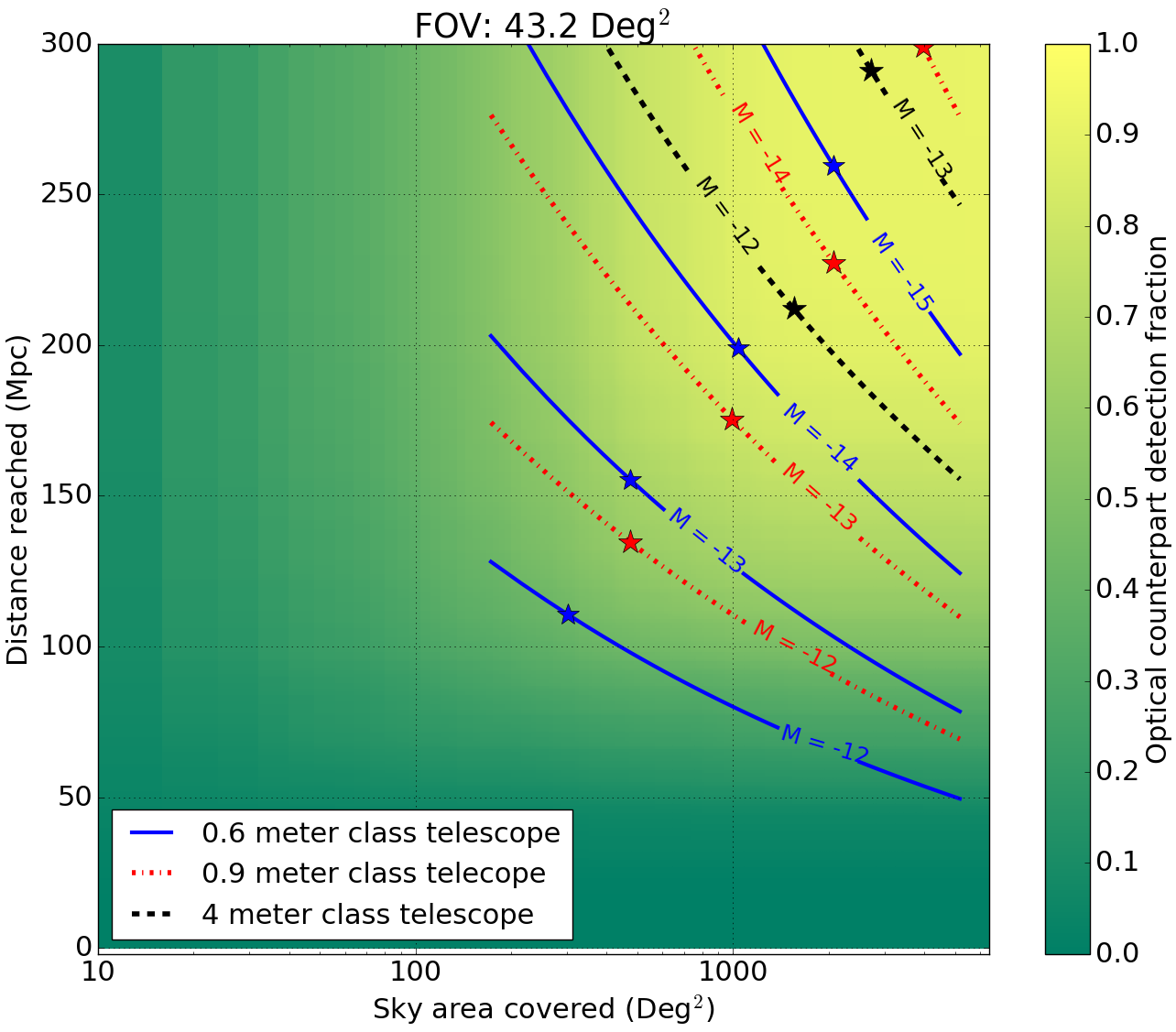}
\caption{Detection fraction as a function of sky area covered and the distance reached in two hours is shown for nine different types of telescopes. Due to uncertainty in kilonova light-curve models we show the detection percentage for four kilonova absolute magnitude cases, $M =-12$, $-13, -14 $ and $-15$. In the $M=-14$ and $-15$ cases for the 4.0 meter class telescope all sources are detectable, hence not shown. }
\label{fig:DetFrac_dist_area}
\end{figure}

However, there is one caveat to this geometrical argument given above, namely here we have assumed a uniform volume distribution of sources. While this is a reasonable assumption for all sources in the universe, the distribution of the number of binary neutron stars from which gravitational waves are detectable by LIGO and Virgo are not uniform in volume. The strength of the gravitational wave signal depends strongly on the inclination angle of the binary with face-on systems being stronger emitters than edge-on systems. This introduces a bias in our detectability, namely we are more likely to detect face-on system at larger distances \citep{0004-637X-767-2-124}. Thus the distribution of the detectable sources by LIGO and Virgo will not scale as $r^3$. This is evident from Fig. \ref{fig:DistanceDistribution} where we show the histogram of the detected events from the 2016 scenario study. 
This implies that the apparent benefit that we saw in Fig. \ref{fig:RatioOfInaccessible} from observing deeper at the cost of covering less of the localization region would be less profound (if any). To check this we conducted the following study. For the nine different telescopes with FOVs $= 2.7, 10.8$ and 43.2 deg$^2$ and apertures $= 0.6, 0.9$ and 4.0 meters, we determine the distance that can be reached as a function of the sky-area covered if we have two hours of observation time at our disposal. The detectability of the source depends on the intrinsic brightness of the sources just as they depend on the telescope's FOV and aperture size. However, currently the kilonova light-curve models are not very well constrained. Therefore, we conducted this analysis for four models with absolute magnitudes $M = -12, -13, -14$ and $-15$ \citep{2041-8205-736-1-L21, 0004-637X-775-2-113}. These values are believed to capture the range of kilonova brightness within reasonable accuracy. In Fig.  \ref{fig:DetFrac_dist_area} we show the result of this analysis. The fraction of the optical counterpart that can be detected from the 2016 scenario is shown in the color scale.
Firstly, note that there is no detectability maximum in the observed sky-area (on any slice along the x-axis). This is contrary to what we observed for the case of uniform distribution of sources. Secondly, it is obvious from Fig. \ref{fig:DetFrac_dist_area} that for smaller observed sky-area ($\lesssim$ 150 deg$^2$) there is virtually no benefit from increasing the depth of the observation beyond $\sim 150$ Mpc. 
An observer will be constrained by time and will not be able cover an arbitrarily large sky-localization area to an arbitrarily high depth. Thus an observer would typically fall on the curves like the ones shown in blue-solid, red-dotted and black-dashed lines in Fig. \ref{fig:DetFrac_dist_area} for the  0.6m, 0.9m and 4.0m class telescopes respectively. If an observer wants to cover greater fraction of the localization region, then the observation must be carried out towards the right-end of these curves, while if the observer intends to observe at greater depth then the observation should be conducted at the left-end of the curves. The background color gives us the corresponding detection probability. The location of the star on the lines indicates the depth and coverage at which maximum detectability of optical counterparts is achieved. From the top panel it is evident that for the smaller FOV telescope it is almost never productive to cover less area in favor of observing deeper unless kilonovae are intrinsically very faint. For an intermediate FOV telescope (middle panel) the benefits of observing deeper at the expense of coverage is absent for very large aperture telescopes. For smaller aperture telescopes there is benefit in observing deeper especially if kilonovae are intrinsically faint. For the very large FOV telescope (bottom panel)  it appears that it is almost always beneficial to observe deeper at the expense of coverage, which is understandable since they can cover most of the source locations within few pointings. The observed variation of detection probability as a function of sky-coverage and depth of search is in qualitative agreement with what was found by \cite{0004-637X-767-2-124} for the advanced LIGO-Virgo design sensitivity.

\section{Conclusion}

The discovery of gravitational-wave (GW150914) from the binary black hole has opened a new window in transient astronomy. We expect to detect GW from compact binary systems involving neutron star(s) in the coming years as the LIGO detectors improve their sensitivity and new detectors come online worldwide. A scenario study for the 2015-16 era performed for the binary neutron star systems showcased what we can expect during the initial years in GW localization. This work uses the results of that study to investigate how well we would be able to cover these localizations on the sky using wide FOV optical telescopes. We examined the performance of the coverage for various different types of telescopes with square FOVs. We compared two different ways to tile up the sky to facilitate the observation of the GW sky-localization. The most obvious and simple-minded approach would be to figure out the smallest area of a given confidence interval ($90\%$, $95\%$ etc) on the sky and cover that using the telescopes. We showed in our work that due to the discreteness of the FOV of the optical telescopes, a ranked-tile strategy leads to a better performance as far as number of tiles required is concerned. In this method, we first generate a grid that covers the entire sky. Each grid element (which we call a tile) in this grid is of the size of the FOV of the telescope. Next, instead of finding the GW localization contour we compute the localization probability in each tile. We rank these tiles based on their localization probability values and select the top tiles (ranked-tiles) that cumulatively constitute the required confidence interval. We found in our study that ranked-tiling makes the optimization of the location of the tiles irrelevant. It ensures that the observer can use a fixed grid of tiles, making it more suitable for image subtraction. We compared the performance of the two methods of tiling up the sky for observation for various FOVs and different confidence intervals. Larger FOV telescopes and observations conducted over smaller confidence interval regions showed greater benefits from using the ranked-tiles. The reduced search sky-area required to reach any confidence interval also implies a reduction in the number of false positives. The fact that for an equal number of tiles the ranked-tiles accommodates a greater localization percentage indicates that in an actual search for optical counterpart of GW events, where (in most cases) observers will be constrained by the number of telescope pointings, adopting ranked-tiling strategy will give the observers a better chance of covering the true sky-position of the event.

We investigated the performance of distributed FOV arrays of wide field of view telescope ($\geq$ 1.0 deg$^2$) with that of the traditional monolithic FOV telescopes in scanning of the simulated gravitational wave sky-localization regions. Our studies showed clear benefit from using such arrays with maximum impact being at search areas $\sim 100$ deg$^2$. The distributed FOV arrays need not be a single facility containing an array of identical telescopes. It could very well be multiple wide FOV optical telescopes around the world with diverse FOVs operating in a joint fashion. Non local telescopes in such arrays will have greater sky coverage which could be extremely beneficial in the initial years given the size and structures of the expected GW sky-localizations.

Finally, we studied the effects of depth of observation. Here we analyzed the detectability of the sources using nine different types of telescopes of various FOVs and aperture sizes, each for four different kilonova brightnesses. Our investigation shows that for smaller FOV  telescopes there is no advantage in sacrificing coverage of the sky-localization area to observe deeper unless kilonovae are intrinsically extremely faint. Larger FOV telescopes ($>10$ deg$^2$) can afford to observe deeper by increasing their integration time.

\begin{acknowledgements}
The authors would like to thank Leo Singer, for his meticulous review and suggestions for the work and the contents of the paper. SB and SG were supported by the research programme of the Foundation for Fundamental Research on Matter (FOM), which is partially supported by the Netherlands Organisation for Scientific Research (NWO). SB and PJG acknowledge the Aspen Center for Physics, which is supported by National Science Foundation grant PHY-1066293, where part of this work was performed.
\end{acknowledgements}

\bibliographystyle{aa}
\bibliography{References}

\appendix
\section{Ranked-Tiling requires less number of tiles}
\label{app:proof}
Let us denote the set of all tiles in the grid sorted in descending order as $\{\Theta: \Theta_i > \Theta_{i+1}\,\,\forall \,i\}$, where $\Theta_i$  is the $i$th tile of the sorted grid. The set of all tiles that are required to cover the $A_{95}$ region $(T)$ is a subset of $\Theta$. We define the elements of $T$ as 
\begin{equation} 
\begin{aligned}
T = \{T_1, T_2,..., T_{N_0}\}, \,.
\end{aligned}
\end{equation}
Thus we can write
\begin{equation} 
\begin{aligned}
\sum_{k=1}^{N_0} \Theta_k \geq  \sum_{l=1}^{N_0} T_l\,,
\end{aligned}
\end{equation}
where the subscript $k$ indexes the set of tiles denoted by $\Theta$ and $l$ indexes set of tiles denoted by $T$. The equal sign is for the trivial (and extreme) case that the set $T$ happens to be the highest $N_0$ elements of $\Theta$. If we denote the quantity in the left hand side of the inequality as $\mathcal{C}$ and the one in the right as $C$ then $\exists \,c \in [C, \mathcal{C}]$ that satisfies
\begin{equation} 
\begin{aligned}
c = \sum_{k=1}^{N_1} \Theta_k\,.
\end{aligned}
\end{equation}
If the smallest value of $c$ that satisfies this happens to be $\mathcal{C}$ then $N_1 = N_0$ for any other values of $c < \mathcal{C}$, we get $N_1 < N_0$, i.e, the number of ranked-tiles needed to reach up to a required localization probability, $N_1$, is less than the number of tiles required to cover a given smallest confidence contour.

\end{document}